\documentclass[12pt]{article}
\usepackage{color}
\usepackage[latin1]{inputenc}
\usepackage[T1]{fontenc}
\usepackage[english]{babel}
\usepackage{amsfonts}
\usepackage{amssymb}
\usepackage{graphicx}
\usepackage{dsfont}
\usepackage{epsfig}
\usepackage{ae}
\usepackage{hyperref}
\usepackage{harvard}
\usepackage{amsmath}
\usepackage{multirow}
\usepackage{subfigure}
\usepackage{tabularx}
\usepackage{color}
\usepackage{amsthm}
\usepackage{appendix}

\newcommand{\E}{\mathds{E}}

\newtheorem{proposition}{Proposition}[section]

\newtheorem{re}{Remark}[section]

\newcommand{\be}{\begin{eqnarray}}
\newcommand{\ee}{\end{eqnarray}}
\newcommand{\by}{\begin{eqnarray*}}
\newcommand{\ey}{\end{eqnarray*}}
\newcommand{\bt}{\begin{theo}}
\newcommand{\et}{\end{theo}}
\newcommand{\bl}{\begin{lem}}
\newcommand{\el}{\end{lem}}
\newtheorem*{lem1}{Lemma A.1}

\newtheorem*{lem2}{Lemma A.2}

\newtheorem*{lem3}{Lemma A.3}

\newcommand{\bc}{\begin{co}}
\newcommand{\ec}{\end{co}}
\newcommand{\eex}{\end{exa}\vspace{-3mm}}
\newcommand{\br}{\begin{re}}
\newcommand{\er}{\end{re}\vspace{-3mm}}
\renewcommand{\geq}{\geqslant}
\renewcommand{\leq}{\leqslant}

\begin{document}
\bibliographystyle{econometrica}
\citationstyle{agsm}

\title{On ``A General Framework for Pricing Asian Options Under Markov Processes"}
\author{   \textsc{Zhenyu Cui} \footnote{Financial Engineering division, School of Systems and Enterprises, Stevens Institute of Technology. Email: \href{mailto:zcui6@stevens.edu}{zcui6@stevens.edu} } \ \quad \ \textsc{Chihoon Lee}\footnote{School of Business, Stevens Institute of Technology. Email: \href{mailto:clee4@stevens.edu}{clee4@stevens.edu}.} \quad \ \textsc{Yanchu Liu}\footnote{Department of Finance, Lingnan (University) College, Sun Yat-Sen University. Email: \href{mailto:liuych26@mail.sysu.edu.cn }{liuych26@mail.sysu.edu.cn } } }

\maketitle

\begin{abstract}
\cite{CSK15} [Cai, N., Y. Song, S. Kou (2015) A general framework for pricing Asian options under Markov processes. \textit{Oper. Res}. 63(3): 540-554] made a breakthrough by proposing a general framework for pricing both discretely and continuously monitored Asian options under one-dimensional Markov processes. In this note, under the setting of continuous-time Markov chain (CTMC), we explicitly carry out the inverse $\mathcal{Z}-$transform and the inverse Laplace transform respectively for the discretely and the continuously monitored cases. The resulting explicit \textit{single} Laplace transforms improve their Theorem 2, p.543, and numerical studies demonstrate the gain in efficiency.
\end{abstract}
\vspace{0.5cm}

\newpage

\section{Introduction\label{s1}}
Asian options are popular path-dependent options actively traded in the financial markets, yet their valuation is challenging and has attracted a significant amount of interest in the literature (\cite{CK12},  \cite{CLS14},  \cite{F99},  \cite{L04}). A breakthrough was achieved in  \cite{CSK15}, where the authors proposed a general framework for pricing both discretely and continuously monitored Asian options under one-dimensional Markov processes through a novel and elegant functional equation approach. They employed results from  \cite{MP13}, a weak approximation scheme from the continuous-time Markov chain (CTMC) to the Markov process, and explicitly solved the functional equations in the CTMC case. The resulting approximation is shown to yield an efficient valuation of Asian option prices. The analytical solutions provided in their paper are given as a $\mathcal{Z}-$Laplace transform for the discrete case, and a two-dimensional Laplace transform for the continuous case (Theorem 2, p.543 of \cite{CSK15}).

In this note, under the setting of CTMC, we explicitly carry out the inverse $\mathcal{Z}-$transform and the inverse Laplace transform, and hence obtain explicit \textit{single} Laplace transforms for both discretely and continuously monitored Asian options. This improves the \textit{double} transforms in Theorem 2 of  \cite{CSK15}.  As a result, the theoretical complexity and computational efficiency are improved accordingly.

\section{Main Result\label{s3}}
Given the filtered probability space $(\Omega, \mathcal{F}, \{\mathcal{F}_t\}_{t\geq 0}, \mathbb{P})$, and assume that we work under the risk-neutral measure $\mathbb{P}$. Denote $r$ the risk-free interest rate, and assume that the dividend rate is $0$.

One can construct a suitable CTMC to approximate a general one-dimensional Markov process (the stock price); see \cite{MP13} or p.544 of  \cite{CSK15}. Thus in the following, we shall restrict our discussions to CTMC.

We consider a non-negative CTMC, $\{X_t\}_{t\geq 0}$ with finite state space $\lbrace x_1,\ldots,x_N\rbrace$, whose transition probability matrix is ${\bf P}(t)=(p_{ij}(t))_{N\times N}$, where $p_{ij}(t)=\mathbb{P}(X_{t+u}=x_j \mid X_u=x_i), 1\leq i,j\leq N, t, u\geq 0$. Its transition rate matrix is ${\bf G}=(q_{ij})_{N\times N}$, where  $q_{ij}=p_{ij}^{\prime}(0),$ $1\leq i,j\leq N$. Define ${\bf x}=(x_1,\ldots,x_N)^{T}$, and let ${\bf I}$ denote the identity matrix. We use ${\bf 1}$ to denote an $N\times 1$ column vector with all entries equal to $1$.
Let ${\bf D}=(d_{ij})_{N\times N}$ be a diagonal matrix with $d_{jj}=x_j$, $j=1,\ldots,N$. 
We use $\E^{ x}[\,\cdot\,]$ to denote the expectation conditional on $X_0=x$.

Consider the following payoff functions studied in equation (1), p.541 of \cite{CSK15}:
\begin{align}
v_c(t,k; x)&=\E^{ x}[(A_t-k)^+], \quad \quad \quad v_d(n,k; x)=\E^{ x}[(B_n-k)^+],\notag
\end{align}
where $A_t:=\int_0^t X_u du$ and $B_n:=\sum\limits_{i=0}^n X_{t_i}$.
Denoting $T$ the maturity and $K$ the strike price,  the price of the continuously monitored Asian call option $V_c(T,K; x)$ at time $0$ is given by $(e^{-rT}/T)v_c(T,TK; x)$. Similarly, the price of the discretely monitored Asian call option $V_d(n,K;x)$ at time $0$ is given by $(e^{-rT}/(n+1))v_d(n,(n+1)K;x)$. Henceforth, $\Delta=T/n$. The following is our main result, which improves Theorem 2, p.543 of  \cite{CSK15}.
Define $$v_c(n,k;{\bf x}):=(v_c(n,k;x_1),\ldots, v_c(n,k;x_N))^T, \quad  v_d(n,k;{\bf x}):=(v_d(n,k;x_1),\ldots, v_d(n,k;x_N))^T.$$

\begin{proposition}\label{main} (Single Laplace transforms for fixed strike Asian options)

(i)(Discretely monitored Asian options)

Let $g_d(n,\theta;{\bf x}):=\int_0^{\infty} e^{-\theta k} v_d(n,k;{\bf x})dk$, then for any complex $\theta$ such that $Re(\theta)>0$, we have
\begin{align}
g_d(n,\theta;{\bf x})&=\frac{1}{\theta^2}(e^{-\theta {\bf D}}{\bf P}(\Delta))^n e^{-\theta {\bf D}} {\bf 1}-\frac{1}{\theta^2}{\bf 1}+\frac{{\bf x}}{\theta}\frac{1-e^{(n+1)r\Delta}}{1-e^{r\Delta}}.\label{g_d1}
\end{align}

(ii) (Continuously monitored Asian options)

Let $g_c(t, \theta; {\bf x}):=\int_0^{\infty} e^{-\theta k} v_c(t,k;{\bf x})dk$, then for any complex $\theta$ such that $Re(\theta)>0$, we have
\begin{align}
g_c(t, \theta; {\bf x})&=\frac{1}{\theta^2}e^{ ({\bf G}-\theta {\bf D}) t } {\bf 1}-\frac{1}{\theta^2}{\bf 1}+\frac{{\bf x}}{r\theta}(e^{rt}-1).\label{contfixed}
\end{align}
\end{proposition}

\textbf{Proof of Proposition \ref{main}}.

(i) Let $L_d(z,\theta; {\bf x}):=\sum\limits_{n=0}^{\infty}z^n \int_0^{\infty}e^{-\theta k} v_d(n,k;{\bf x})dk=\sum\limits_{n=0}^{\infty}z^n g_d(n,\theta;{\bf x})$.   From Proposition 1(i) and Theorem 2(i) of \cite{CSK15}, we have
\begin{align}
L_d(z,\theta;{\bf x})&=\frac{1}{\theta^2}(e^{\theta {\bf D}}-z {\bf P}(\Delta))^{-1} {\bf 1}-\frac{1}{\theta^2(1-z)}{\bf 1}+\frac{{\bf x}}{\theta(1-z)(1-z e^{r\Delta})}.\label{z}
\end{align}

From definition, we observe that $g_d(n,\theta;{\bf x})$ can be treated as the coefficient of $z^n$ in the power series expansion of $L_d(z,\theta;{\bf x})$ with respect to the transform variable $z$. This motivates us to expand the right hand side of \eqref{z} into a power series of $z$.

Similar as in the proof of Theorem 2 in  \cite{CSK15}, we can show that $e^{\theta {\bf D}}-z {\bf P}(\Delta)$ is strictly diagonally dominant, and by the L\'{e}vy-Desplanques theorem (Corollary $5.6.17$ of \cite{HJ85}), we have that $e^{\theta {\bf D}}-z {\bf P}(\Delta)$ is invertible.
From Corollary $5.6.16$ of \cite{HJ85}, if there is a matrix norm $|| \cdot||$ (without loss of generality, we can take the maximum norm, i.e., $|| {\bf A} ||=\max \{ |a_{ij}| \})$ such that $|| {\bf I}-{\bf A}||<1$, then we have ${\bf A}^{-1}=\sum_{k=0}^{\infty} ({\bf I}-{\bf A})^k$.

 In the following, we assume that the transform variable satisfies $|z|<\min\{1, e^{-r\Delta}, 1/|| (e^{\theta {\bf D}})^{-1}{\bf P}(\Delta) ||\}$, so that the power  series expansions with respect to $z$ are well-defined. We have
\begin{align}
&(e^{\theta {\bf D}}-z {\bf P}(\Delta))^{-1} {\bf 1}=(e^{\theta {\bf D}}({\bf I}-z (e^{\theta {\bf D}})^{-1}{\bf P}(\Delta)))^{-1} {\bf 1}\notag\\
\quad & =({\bf I}-z (e^{\theta {\bf D}})^{-1}{\bf P}(\Delta))^{-1} (e^{\theta {\bf D}})^{-1} {\bf 1}\notag\\
\quad &=\left({\bf I}+ z (e^{\theta {\bf D}})^{-1}{\bf P}(\Delta) +z^2 ((e^{\theta {\bf D}})^{-1}{\bf P}(\Delta))^2+\cdots+z^n ((e^{\theta {\bf D}})^{-1}{\bf P}(\Delta))^n+\cdots\right)(e^{\theta {\bf D}})^{-1} {\bf 1}\notag\\
\quad &=(e^{\theta {\bf D}})^{-1} {\bf 1} +z (e^{\theta {\bf D}})^{-1}{\bf P}(\Delta)(e^{\theta {\bf D}})^{-1} {\bf 1}+\cdots+z^n ((e^{\theta {\bf D}})^{-1}{\bf P}(\Delta))^n (e^{\theta {\bf D}})^{-1} {\bf 1}+\cdots
\label{ek1}
\end{align}

The coefficient of $z^n$ in \eqref{ek1} is $((e^{\theta {\bf D}})^{-1}{\bf P}(\Delta))^n (e^{\theta {\bf D}})^{-1} {\bf 1}$. 
Expand the remaining parts in \eqref{z} as
\begin{align}
-\frac{1}{\theta^2(1-z)}{\bf 1}&=-\frac{1}{\theta^2}{\bf 1}\sum\limits_{i=0}^{\infty} z^i ,\label{e2}
\end{align}
and
\begin{align}
\frac{{\bf x}}{\theta(1-z)(1-z e^{r\Delta})}&=\frac{{\bf x}}{\theta} \sum\limits_{i=0}^{\infty} z^i \times \sum\limits_{j=0}^{\infty} z^j e^{jr\Delta}.\label{e3}
\end{align}

It is clear that the coefficient of $z^n$ in \eqref{e2} is $-\frac{1}{\theta^2}{\bf 1}$, and the coefficient of $z^n$ in \eqref{e3} is $\frac{{\bf x}}{\theta}\frac{1-e^{(n+1)r\Delta}}{1-e^{r\Delta}}$.

Thus the inverse $\mathcal{Z}-$transform of $L_d(z,\theta;x)$ is given by
\begin{align}
g_d(n,\theta;{\bf x})&=\mathcal{Z}^{-1}\left(L_d(z,\theta;x)\right)=\frac{1}{\theta^2}((e^{\theta {\bf D}})^{-1}{\bf P}(\Delta))^n (e^{\theta {\bf D}})^{-1} {\bf 1}-\frac{1}{\theta^2}{\bf 1}+\frac{{\bf x}}{\theta}\frac{1-e^{(n+1)r\Delta}}{1-e^{r\Delta}},\label{final1}
\end{align}
and, together with $(e^{\theta {\bf D}})^{-1}=e^{-\theta {\bf D}}$, it completes the proof of part (i).

(ii) Let $L_c(\mu,\theta;{\bf x}):=\int_0^{\infty}e^{-\mu t} \int_0^{\infty} e^{-\theta k} v_c(t,k;{\bf x})dkdt=\int_0^{\infty}e^{-\mu t} g_c(t, \theta; {\bf x}) dt$.  From Proposition 1(ii) and Theorem 2(ii) of  \cite{CSK15}, we have
\begin{align}
L_c(\mu,\theta;{\bf x})&=\frac{1}{\theta^2}m(\mu,\theta; {\bf x})-\frac{1}{\theta^2\mu}{\bf 1}+\frac{{\bf x}}{\theta\mu(\mu-r)},\label{Lc}
\end{align}
where $m(\mu,\theta; {\bf x}):=(\theta {\bf D}+\mu {\bf I}-{\bf G})^{-1} {\bf 1}$.

It can be shown that $\theta {\bf D}+\mu {\bf I}-{\bf G}$ is strictly diagonally dominant, and thus invertible by the L\'{e}vy-Desplanques theorem. We assume $|\mu|>\max\{|| {\bf G}-\theta {\bf D} ||, 0\}$,  so that the following power series expansions with respect to $\mu$ are well-defined.
We obtain
\begin{align}
(\theta {\bf D}+\mu {\bf I}-{\bf G})^{-1} {\bf 1}&=\left(\mu \left({\bf I}-\left(-\frac{\theta}{\mu} {\bf D}+\frac{1}{\mu}{\bf G}\right)\right)\right)^{-1} {\bf 1}\notag\\
&=\frac{1}{\mu} \left({\bf I}-\frac{{\bf G}-\theta {\bf D}}{\mu}\right)^{-1}   {\bf 1}\notag\\
&=\frac{1}{\mu}\left({\bf I}+ \frac{{\bf G}-\theta {\bf D}}{\mu} +\left(\frac{{\bf G}-\theta {\bf D}}{\mu}\right)^2+ \cdots+ \left(\frac{{\bf G}-\theta {\bf D}}{\mu}\right)^n+\cdots\right)  {\bf 1}\notag\\
&=\frac{{\bf 1}}{\mu}+\frac{({\bf G}-\theta {\bf D})  {\bf 1}}{\mu^2}+\frac{({\bf G}-\theta {\bf D})^2   {\bf 1}}{\mu^3}+\cdots+\frac{({\bf G}-\theta {\bf D})^{n}   {\bf 1}}{\mu^{n+1}}+\cdots.\label{ee2}
\end{align}

It is clear from \eqref{ee2} that the inverse Laplace transform of $m(\mu,\theta;{\bf x})$ with respect to $\mu$ is given by
\begin{align}
\mathcal{L}_{\mu}^{-1}(m(\mu,\theta;{\bf x}))&=\sum\limits_{i=0}^{\infty}  \frac{({\bf G}-\theta {\bf D})^i   {\bf 1}}{i!}  t^i=e^{ ({\bf G}-\theta {\bf D}) t }   {\bf 1}.\label{keyinv}
\end{align}

It is easy to identify the inverse Laplace transform with respect to $\mu$ of the two remaining terms in \eqref{Lc} as
\begin{align}
\mathcal{L}_{\mu}^{-1}\left( -\frac{1}{\theta^2 \mu}{\bf 1}\right)&=-\frac{1}{\theta^2}{\bf 1}, \quad \quad \quad  \mathcal{L}_{\mu}^{-1}\left( \frac{{\bf x}}{\theta \mu(\mu-r)}\right)=\frac{{\bf x}}{r\theta}(e^{rt}-1).\label{keyinv2}
\end{align}

From \eqref{keyinv} and \eqref{keyinv2}, we have
\begin{align}
g_c(t, \theta; {\bf x})&=\mathcal{L}_{\mu}^{-1}\left(L_c(\mu,\theta;{\bf x})\right)=\frac{1}{\theta^2}e^{ ({\bf G}-\theta {\bf D}) t }   {\bf 1}-\frac{1}{\theta^2}{\bf 1}+\frac{{\bf x}}{r\theta}(e^{rt}-1).\label{final2}
\end{align}

This completes the proof. $\Box$
\endproof

\begin{re}
We note that it is very difficult, if not impossible, to further carry out Laplace inversion of the function $g_d(n,\theta;{\bf x})$ or $g_c(t, \theta; {\bf x})$ and hence obtain closed-form expressions for the option prices. For $g_d(n,\theta;{\bf x})$, we can easily identify the inverse Laplace transform of $\frac{1}{\theta^2}e^{-\theta {\bf D}}$ and $e^{-\theta {\bf D}}{\bf P}(\Delta)$, respectively, but it is challenging to obtain a tractable Laplace inversion of $(e^{-\theta {\bf D}}{\bf P}(\Delta))^n$. For $g_c(t, \theta; {\bf x})$, the major difficulty for carrying out further Laplace inversion lies in the matrix exponential $e^{ ({\bf G}-\theta {\bf D}) t }$, and the fact that ${\bf G}$ and ${\bf D}$ are in general not commutable.
  Based on these observations, we argue that the results in Proposition \ref{main} may not be further improved in general.
\end{re}

\begin{re}
There are two ways of computing the expression given in \eqref{g_d1}. In the ``forward" way, we first compute the $n$th power of $e^{-\theta {\bf D}}{\bf P}(\Delta)$ and then multiply it by the vector $e^{-\theta {\bf D}} {\bf 1}$. The complexity would be $O(N^3n)$, since the multiplication of two $N-$dimensional square matrices has cost of $O(N^3)$. In the ``backward" way, we multiply the matrix $e^{-\theta {\bf D}}{\bf P}(\Delta)$ by the vector $e^{-\theta {\bf D}} {\bf 1}$ first and the result is again a vector. The complexity of this operation is only $O(N^2)$, because: (1) $e^{-\theta {\bf D}}$ is a diagonal matrix and therefore computing $(e^{\theta {\bf D}})^{-1}{\bf P}(\Delta)$ costs only $O(N)$, and (2) Multiplication of an $N-$dimensional square matrix by a vector costs $O(N^2)$. Repeating this procedure $n$ times, we obtain $((e^{\theta {\bf D}})^{-1}{\bf P}(\Delta))^n (e^{\theta {\bf D}})^{-1} {\bf 1}$ with a total cost of $O(N^2n)$, which reduces the complexity by $O(N)$ times compared with the ``forward" way. Although in the numerical tests conducted in the following section, we did not observe significant differences between the realized running time of these two ways of implementation via Matlab, we believe that the ``backward" way will have some potential benefits in more computationally intensive settings.
\end{re}

\section{Numerical Results\label{s4}}
We follow the two-step procedure proposed in \S 5 (p. 544) of \cite{CSK15} to compute the Asian option prices, except that in the second step we only need to invert the  \textit{single} Laplace transforms given in (\ref{g_d1}) and (\ref{contfixed}). Note that inverting a single Laplace transform is well studied, and we use the equations (4.6) and (6.26) in \cite{AW92}. Numerical results are collected under different models, including the CIR model, the CEV model, the double-exponential jump diffusion (DEJD) model, the Merton's jump diffusion (MJD) model, and the Carr-Geman-Madan-Yor (CGMY) model. Parameter settings for each model involved are exactly the same as those in \S 5 of \cite{CSK15}. All numerical experiments in this note are conducted using Matlab R2014b on a laptop equipped with an Intel Core 2 i7-4510U CPU @ 2.00 GHz 2.60 GHZ and 8 GB of RAM.

The following tables report the Asian option prices computed via our single Laplace transform formulas, and compare them with certain benchmarks and results based on the double transform methods, both of which are directly taken from \cite{CSK15}. Relative errors of our method compared to the benchmarks are also presented. We also record CPU running times of our method.

We observe similar patterns across the tables: (1) The option prices yielded by our single Laplace transform inversions are very close to those from the double transform inversions in \cite{CSK15}, while it takes  less realized running time in all cases. (2) For discretely monitored Asian options, the running time of our method increases with the number of monitoring points, i.e., $n$, but at a  slower rate than that reported in \cite{CSK15}.

\section{Conclusion\label{s6}}

In this note, under the setting of CTMC, we obtain explicit single Laplace transforms for prices of discretely  and  continuously monitored Asian options.  This improves Theorem 2, p.543 of  \cite{CSK15}, and numerical studies demonstrate the gain in efficiency.

\bibliographystyle{ormsv080} 
\bibliography{References}
\newpage

\begin{table}[!htbp]\footnotesize
\caption{\textbf{Asian options under the CIR model.}} 
\label{tab:CIR}
\centering 
\begin{tabular}{c c c c c c c c c}
\hline
    $K$     & Benchmark & Cai et al. & CTMC & Rel. err.($\%$) & Benchmark & Cai et al. & CTMC & Rel. err.($\%$)\\
\hline
& \multicolumn{4}{c}{$n=12$} & \multicolumn{4}{c}{$n=25$} \\
0.90 & 0.21279   & 0.21257 & 0.21300 & 0.10 & 0.21428 & 0.21406 & 0.21449 & 0.10  \\ 
0.95 & 0.18659   & 0.18638 & 0.18674 & 0.08 & 0.18810 & 0.18789 & 0.18823 & 0.07 \\  
1.00 & 0.16282   & 0.16264 & 0.16297 & 0.09 & 0.16432 & 0.16414 & 0.16445 & 0.08  \\ 
1.05 & 0.14140   & 0.14126 & 0.14158 & 0.13 & 0.14287 & 0.14273 & 0.14303 & 0.11 \\ 
1.10 & 0.12223   & 0.12213 & 0.12245 & 0.18 & 0.12365 & 0.12355 & 0.12385 & 0.17 \\ 
& \multicolumn{4}{c}{$n=50$} & \multicolumn{4}{c}{$n=100$} \\
0.90 & 0.21501   & 0.21406 &0.21521  &0.09 & 0.21538 & 0.21515 & 0.21558  &0.09  \\ 
0.95 & 0.18883   & 0.18862 &0.18896  &0.07 & 0.18920 & 0.18899 & 0.18933  &0.07 \\  
1.00 & 0.16505   & 0.16487 &0.16517  &0.07 & 0.16542 & 0.16524 & 0.16554  &0.07   \\ 
1.05 & 0.14359   & 0.14344 &0.14374  &0.10 & 0.14395 & 0.14381 & 0.14410  &0.10 \\ 
1.10 & 0.12434   & 0.12424 &0.12453  &0.15 & 0.12470 & 0.12460 & 0.12489  &0.15 \\ 
& \multicolumn{4}{c}{$n=250$} & \multicolumn{4}{c}{$n=+\infty$} \\
0.90 & 0.21560   & 0.21537 & 0.21581 & 0.10 & 0.21575 & 0.21552 & 0.21592 & 0.08  \\ 
0.95 & 0.18943   & 0.18922 & 0.18956 & 0.07 & 0.18958 & 0.18937 & 0.18976 & 0.09 \\  
1.00 & 0.16565   & 0.16547 & 0.16578 & 0.08 & 0.16580 & 0.16562 & 0.16600 & 0.12  \\ 
1.05 & 0.14418   & 0.14403 & 0.14432 & 0.10 & 0.14433 & 0.14418 & 0.14457 & 0.17 \\ 
1.10 & 0.12492   & 0.12481 & 0.12510 & 0.14 & 0.12506 & 0.12496 & 0.12534 & 0.22 \\
\hline
\end{tabular}
\vspace{0.1cm}
\begin{flushleft}
Note. Pricing Asian options under the CIR model via our single transform CTMC approximation with $N=50$. The parameter settings are the same as those in \cite{FMR08} and also in \S 5.1 of \cite{CSK15}. The columns ``Benchmark" are taken from \cite{FMR08}, whose values are computed from their analytical solutions. The columns ``Cai et al." are taken from the Table 3 of \cite{CSK15}. Results based on our single Laplace transforms are presented in the columns ``CTMC." The columns ``Rel. err.($\%$)" document the relative errors of our method compared with the benchmark values. The realized computing times to output the option price based on our method are about \textbf{0.009}, \textbf{0.011}, \textbf{0.013}, \textbf{0.016}, and \textbf{0.023} seconds for $n$ = 12, 25, 50, 100, and 250, respectively, and about \textbf{0.031} seconds for $n=+\infty$ (i.e., the continuously monitored Asian options).
\end{flushleft}
\end{table}

\newpage

\begin{table}[!htbp]\footnotesize
\caption{\textbf{Asian options under the CEV model.}} 
\label{tab:CEV}
\centering 
\begin{tabular}{c c c c c c c c c}
\hline
& \multicolumn{4}{c}{(I) Discretely monitored Asian options} & \multicolumn{4}{c}{(II) Continuously monitored Asian options} \\
& \multicolumn{4}{c}{under the CEV model ($n=250$)} & \multicolumn{4}{c}{under the CEV model} \\
    $K$     & Benchmark & Cai et al. & CTMC & Rel. err.($\%$) & Benchmark & Cai et al. & CTMC & Rel. err.($\%$)\\
\hline
& \multicolumn{4}{c}{$\beta=0.25$} & \multicolumn{4}{c}{$\beta=0.25$} \\
80  & 21.60167  & 21.60974& 21.60980 & 0.04 & 21.59408(0.00468) & 21.61076 & 21.61093 & 0.08  \\ 
90  & 13.15550  & 13.15548& 13.15551 & 0.00 & 13.15109(0.00425) & 13.15931 & 13.15920 & 0.07 \\  
100 & 6.84034   & 6.82619 & 6.82623  & 0.21 & 6.83859(0.00340)  & 6.83128  & 6.83146  & 0.10  \\ 
110 & 3.07180   & 3.05691 & 3.05697  & 0.48 & 3.07333(0.00239)  & 3.06138  & 3.06136  & 0.39 \\ 
120 & 1.22841   & 1.22497 & 1.22502  & 0.28 & 1.23175(0.00154)  & 1.22762  & 1.22765  & 0.33 \\ 
& \multicolumn{4}{c}{$\beta=-0.25$} & \multicolumn{4}{c}{$\beta=-0.25$} \\
80  & 21.67122  & 21.67979& 21.67979 & 0.04 & 21.66618(0.00464) & 21.68104 & 21.68112 & 0.07  \\ 
90  & 13.26903  & 13.26768& 13.26768 & 0.01 & 13.26741(0.00417) & 13.27147 & 13.27137 & 0.03 \\  
100 & 6.84853   & 6.83407 & 6.83409  & 0.21 & 6.85150(0.00327)  & 6.83920  & 6.83932  & 0.18  \\ 
110 & 2.92962   & 2.91597 & 2.91599  & 0.46 & 2.93166(0.00221)  & 2.92049  & 2.92050  & 0.38 \\ 
120 & 1.04072   & 1.04152 & 1.04154  & 0.08 & 1.04453(0.00131)  & 1.04429  & 1.04420  & 0.03 \\ 
& \multicolumn{4}{c}{$\beta=-0.5$} & \multicolumn{4}{c}{$\beta=-0.5$} \\
80  & 21.71428  & 21.72237& 21.72238 & 0.04 & 21.71118(0.00465) & 21.72370 & 21.72379 & 0.06  \\ 
90  & 13.32877  & 13.32675& 13.32676 & 0.02 & 13.32850(0.00416) & 13.33052 & 13.33044 & 0.01 \\  
100 & 6.85365   & 6.83904 & 6.83906  & 0.21 & 6.85984(0.00324)  & 6.84420  & 6.84429  & 0.23  \\ 
110 & 2.86119   & 2.84823 & 2.84824  & 0.45 & 2.86666(0.00215)  & 2.85276  & 2.85281  & 0.48 \\ 
120 & 0.95542   & 0.95803 & 0.95805  & 0.28 & 0.95995(0.00122)  & 0.96084  & 0.96070  & 0.08 \\ 
\hline
\end{tabular}
\vspace{0.1cm}
\begin{flushleft}
Note. Pricing Asian options under the CEV model via our single transform CTMC approximation with $N=50$. All parameter settings are the same as those in \S 5.2 of \cite{CSK15}. Part (I) compares our results (i.e., the column ``CTMC") with the asymptotic expansion numerical prices in \cite{CLS14} (i.e., the column ``Benchmark") and results via the double transform method in \cite{CSK15} (i.e., the column ``Cai et al."). These two reference columns are taken from Part (I) of Table 4 in \cite{CSK15}. It takes about \textbf{0.024} seconds on average to output one price by our method. Part (II) collects results for continuously monitored Asian options. The benchmark values, taken from Part (II) of Table 4 in \cite{CSK15}, are based on Monte Carlo simulations with numbers in the brackets indicating standard deviations.  The column ``Cai et al.", taken from the same reference, are obtained based on the double transform method in \cite{CSK15}. It takes about \textbf{0.049} seconds on average to output a price via our method.
\end{flushleft}
\end{table}

\newpage

\begin{table}[!htbp]\scriptsize
\caption{\textbf{Asian options under the DEJD model.}} 
\label{tab:DEJD}
\centering 
\begin{tabular}{c c c c c c c c c}
\hline
\multicolumn{9}{c}{(I) Discretely monitored Asian options under the DEJD model} \\
\hline
$n$ &    $K$     & \multicolumn{2}{c}{Benchmark} & \multicolumn{2}{c}{Cai et al.} & CTMC & \multicolumn{2}{c}{Rel. err.($\%$)} \\
\hline
12&90  &\multicolumn{2}{c}{12.71236} & \multicolumn{2}{c}{12.70857} &12.70873& \multicolumn{2}{c}{0.03}  \\ 
  &100 &\multicolumn{2}{c}{5.01712}  & \multicolumn{2}{c}{5.01254}  &5.01263 & \multicolumn{2}{c}{0.09}  \\ 
  &110 &\multicolumn{2}{c}{1.04142}  & \multicolumn{2}{c}{1.03988}  &1.03989 & \multicolumn{2}{c}{0.15}  \\ 
50&90  &\multicolumn{2}{c}{12.74369} & \multicolumn{2}{c}{12.74016} &12.74025& \multicolumn{2}{c}{0.03}  \\ 
  &100 &\multicolumn{2}{c}{5.05809}  & \multicolumn{2}{c}{5.05358}  &5.05371 & \multicolumn{2}{c}{0.09}  \\ 
  &110 &\multicolumn{2}{c}{1.06878}  & \multicolumn{2}{c}{1.06725}  &1.06725 & \multicolumn{2}{c}{0.14}  \\ 
250&90 &\multicolumn{2}{c}{12.75241} & \multicolumn{2}{c}{12.74875} &12.74881& \multicolumn{2}{c}{0.03}  \\ 
   &100&\multicolumn{2}{c}{5.06949}  & \multicolumn{2}{c}{5.06491}  &5.06504 & \multicolumn{2}{c}{0.09}  \\ 
   &110&\multicolumn{2}{c}{1.07646}  & \multicolumn{2}{c}{1.07489}  &1.07489 & \multicolumn{2}{c}{0.15}  \\ 
\hline
\multicolumn{9}{c}{(II) Continuously monitored Asian options under the DEJD model} \\
\hline
$K$ &  Benchmark  & Cai et al. & CTMC & Rel. err.($\%$) & Benchmark  & Cai et al. & CTMC & Rel. err.($\%$)\\
\hline
& \multicolumn{4}{c}{$\sigma=0.05$} & \multicolumn{4}{c}{$\sigma=0.1$} \\
90  & 13.47952  & 13.46823& 13.47752 & 0.01  & 13.55964 & 13.56418 & 13.56389 & 0.03  \\ 
95  & 9.16588   & 9.18472 & 9.16582  &0.0007 & 9.41962  & 9.42931  & 9.42470  & 0.05 \\  
100 & 5.38761   & 5.37399 & 5.38772  &0.002  & 5.91537  & 5.91365  & 5.91780  & 0.04  \\ 
105 & 2.72681   & 2.71628 & 2.72530  & 0.06  & 3.35071  & 3.34830  & 3.35143  & 0.02 \\ 
110 & 1.28264   & 1.30224 & 1.28198  & 0.05  & 1.74896  & 1.75431  & 1.74943  & 0.03 \\ 
& \multicolumn{4}{c}{$\sigma=0.2$} & \multicolumn{4}{c}{$\sigma=0.3$} \\
80  & 14.17380  & 14.17589& 14.17568 & 0.01 & 15.33688 & 15.33545 & 15.33575 & 0.007  \\ 
90  & 10.53795  & 10.53824& 10.53807 &0.001 & 12.10723 & 12.10414 & 12.10441 & 0.01 \\  
100 & 7.48805   & 7.48621 & 7.48648  & 0.02 & 9.35336  & 9.34883  & 9.34914  & 0.02  \\ 
110 & 5.09001   & 5.08708 & 5.08736  & 0.05 & 7.08059  & 7.07520  & 7.07551  & 0.07 \\ 
120 & 3.32061   & 3.31802 & 3.31789  & 0.08 & 5.26109  & 5.25561  & 5.25589  & 0.10 \\ 
& \multicolumn{4}{c}{$\sigma=0.4$} & \multicolumn{4}{c}{$\sigma=0.5$} \\
80  & 16.81490  & 16.81130& 16.81958 & 0.03 & 18.46259 & 18.45288 & 18.46148 & 0.006  \\ 
90  & 13.87995  & 13.87460& 13.88190 & 0.01 & 15.75006 & 15.73859 & 15.74575 & 0.03 \\  
100 & 11.33257  & 11.32581& 11.33275 &0.002 & 13.36027 & 13.34737 & 13.35386 & 0.05  \\ 
110 & 9.16131   & 9.15366 & 9.16048  &0.009 & 11.27716 & 11.26330 & 11.26950 & 0.07 \\ 
120 & 7.34063   & 7.33266 & 7.33944  & 0.02 & 9.47826  & 9.46389  & 9.47003  & 0.09 \\ 
\hline
\end{tabular}
\vspace{0.1cm}
\begin{flushleft}
Note. Pricing Asian options under the DEJD model via our single transform CTMC approximation with $N=50$. All parameter settings are the same as those in \S 5.3 of \cite{CSK15}. Part (I) compares our results (i.e., the column ``CTMC") with numerical prices obtained by the recursive algorithm in \cite{FM08} (i.e., the column ``Benchmark") and results via the double transform method in \cite{CSK15} (i.e., the column ``Cai et al."). These two reference columns are taken from the Part (I) of Table 5 in \cite{CSK15}. The CPU times to output the option prices based on our method are about \textbf{0.015}, \textbf{0.018}, and \textbf{0.036} seconds for $n$ = 12, 50, and 250, respectively. Part (II) collects results for continuously monitored Asian options, where the benchmark values are from \cite{CK12}. The column ``Cai et al.",  taken from Part (II) of Table 5 in \cite{CSK15}, is based on the double transform method. It takes about \textbf{0.21} seconds to compute one price via our method for the continuously monitored Asian options.
\end{flushleft}
\end{table}


\begin{table}[!htbp]\footnotesize
\caption{\textbf{Asian options under the MJD model.}} 
\label{tab:MJD}
\centering 
\begin{tabular}{c c c c c c}
\hline
\multicolumn{6}{c}{(I) Discretely monitored Asian options under the MJD model} \\
\hline
$n$ &    $K$     & Benchmark & Cai et al. & CTMC & Rel. err.($\%$) \\
\hline
12&90  & 12.71066 & 12.70620 &12.70636& 0.03  \\ 
  &100 & 5.01127  & 5.00539  & 5.00546& 0.12  \\ 
  &110 & 1.05162  & 1.04941  &1.04940 & 0.21  \\ 
50&90  & 12.74093 & 12.73659 &12.73665& 0.03 \\ 
  &100 & 5.05246  & 5.04654  &5.04667 & 0.11 \\
  &110 & 1.07959  & 1.07736  &1.07733 & 0.21 \\
250&90 & 12.74917 & 12.74485 &12.74490& 0.03  \\
   &100& 5.06381  & 5.05790  &5.05803 & 0.11  \\
   &110& 1.08740  & 1.08515  &1.08512 & 0.21 \\
\hline
\multicolumn{6}{c}{(II) Continuously monitored Asian options under the MJD model} \\
\hline
$K$     & \multicolumn{2}{c}{Benchmark} & Cai et al. & CTMC & Rel. err.($\%$) \\
\hline
90  & \multicolumn{2}{c}{12.74857(0.00371)} & 12.74705 & 12.74699 & 0.01  \\
100 & \multicolumn{2}{c}{5.05974(0.00399)}  & 5.05740  & 5.06095  & 0.02  \\
110 & \multicolumn{2}{c}{1.08413(0.00280)}  & 1.09235  & 1.08712  & 0.28  \\
\hline

\end{tabular}
\vspace{0.1cm}
\begin{flushleft}
Note. Pricing Asian options under the MJD model via our single transform CTMC approximation with $N=50$. All parameter settings are the same as those in \S 5.4 of \cite{CSK15}. Part (I) compares our results (i.e., the column ``CTMC") with numerical prices obtained by the recursive algorithm in \cite{FM08} (i.e., the column ``Benchmark") and results via the double transform method in \cite{CSK15} (i.e., the column ``Cai et al."). These two reference columns are taken from the Part (I) of Table 6 in \cite{CSK15}. The CPU times to output the option price based on our method are about \textbf{0.008}, \textbf{0.011}, and \textbf{0.025} seconds for $n$ = 12, 50, and 250, respectively. Part (II) collects results for continuously monitored Asian options.  The benchmark values, taken from Part (II) of Table 6 in \cite{CSK15}, are based on Monte Carlo simulations with numbers in the brackets indicating standard deviations. The column ``Cai et al.", taken from the same reference, is based on the double transform method. It takes about \textbf{0.047} seconds to compute one price via our method for the continuously monitored Asian options.
\end{flushleft}
\end{table}

\newpage

\begin{table}[!htbp]\footnotesize
\caption{\textbf{Asian options under the CGMY model.}} 
\label{tab:CGMY}
\centering 
\begin{tabular}{c c c c c c}
\hline
\multicolumn{6}{c}{(I) Discretely monitored Asian options under the CGMY model} \\
\hline
$n$ &    $K$     & Benchmark & Cai et al. & CTMC & Rel. err.($\%$) \\
\hline
12&90  & 12.70625 & 12.70406 &12.70318& 0.02  \\
  &100 & 5.03492  & 5.02551  & 5.02612& 0.17  \\
  &110 & 1.02115  & 1.01464  &1.01304 & 0.79  \\
50&90  & 12.73854 & 12.73745 &12.73644& 0.02 \\
  &100 & 5.07570  & 5.06651  &5.06716 & 0.17 \\
  &110 & 1.04674  & 1.04012  &1.03854 & 0.78 \\
250&90 & 12.74737 & 12.74653 &12.74549& 0.01  \\
   &100& 5.08694  & 5.07783  &5.07849 & 0.17  \\
   &110& 1.05389  & 1.04725  &1.04567 & 0.78 \\
\hline
\multicolumn{6}{c}{(II) Continuously monitored Asian options under the CGMY model} \\
\hline
$K$     & \multicolumn{2}{c}{Benchmark} & Cai et al. & CTMC & Rel. err.($\%$) \\
\hline
90  & \multicolumn{2}{c}{12.74788(0.00396)} & 12.74689 & 12.74780 & 0.0006  \\
100 & \multicolumn{2}{c}{5.08865(0.00405)}  & 5.08019  & 5.08138  & 0.14  \\
110 & \multicolumn{2}{c}{1.05810(0.00280)}  & 1.06028  & 1.05751  & 0.06  \\  
\hline

\end{tabular}
\vspace{0.1cm}
\begin{flushleft}
Note. Pricing Asian options under the CGMY model via our single transform CTMC approximation with $N=50$. All parameter settings are the same as those in \S 5.6 of \cite{CSK15}. Part (I) compares our results (i.e., the column ``CTMC") with numerical prices obtained by the recursive algorithm in \cite{FM08} (i.e., the column ``Benchmark") and results via the double transform method in \cite{CSK15} (i.e., the column ``Cai et al."). These two reference columns are taken from the Part (I) of Table 9 in \cite{CSK15}. The CPU times to output the option price based on our method are about \textbf{0.015}, \textbf{0.018}, and \textbf{0.025} seconds for $n$ = 12, 50, and 250, respectively. Part (II) collects results for continuously monitored Asian options. The benchmark values, taken from Part (II) of Table 9 in \cite{CSK15}, are based on Monte Carlo simulations with numbers in the brackets indicating standard deviations. The column ``Cai et al.", taken from the same reference, is based on the double transform method.
 It takes about \textbf{0.092} seconds to compute one price via our method for the continuously monitored Asian options.
\end{flushleft}
\end{table}
\end{document}